\documentstyle[preprint,aps]{revtex}

\begin{document}
\draft
\title{Strong correlation effects in the doped Hubbard model
       in infinite dimensions}
\author{Hae-Young Kee and Jongbae Hong}
\address{Department of Physics Education, Seoul National University,
Seoul 151-742, Korea}
\maketitle

\begin{abstract}
The density of states and the optical conductivity of the doped Hubbard model
on a Bethe lattice with infinite connectivities have been studied
using an analytic variant of the Lanczos continued fraction method.
The spectral weight of the gap states and the position of the chemical
potential upon hole or electron doping have been studied. We argue that the
strong correlation effects such as gap states and midinfrared band shown in
two dimensions also appear in infinite dimensions.
\end{abstract}

The undoped parent compounds of high-$T_c$ superconductors are
known to have the charge-transfer gap between the upper Hubbard band of
Cu 3d character and the valence band of O 2p character.\cite{za} This is
mainly due to the strong correlation effects of the 3d electrons on the Cu
sites. Upon doping, the system undergoes insulator-metal transition with
creating new states (gap states) around Fermi energy inside the
undoped insulator gap.\cite{dago} Recent experiments investigating the
electronic structure of high-$T_c$ superconductors to understand the normal
state properties have been performed by x-ray photoemission spectroscopy
(PES)\cite{zx,all} on one hand and x-ray
absorption spectroscopy (XAS)\cite{che} on the other hand. Both
clarify the creation of gap states inside the insulator gap upon doping.
The occurrence of the gap states has also been supported by the
mid-infrared (MIR) band in the optical conductivity.\cite{uch}
There is, however, a controversy in the behavior of the chemical potential
between the two measurements.\cite{dago} PES experiments on hole doped $La_{2-
x}Sr_xCuO_{4+\delta} $\cite{zx} and electron doped
$Nd_{2-x}Ce_xCuO_4$\cite{all} claim that the chemical potential
remains at nearly the same position, in the gap states even though the
doping character changes. On the other hand, XAS experiments, which
examine unoccupied states, on the same materials show two prepeaks which
reflect the chemical potential jumps across the undoped insulator gap by
changing the doping character instead of remaining at the same position as
shown in PES experiments. This controversy has not been clarified yet.

In the theoretical side, the simple one-band Hubbard model may be chosen
as an effective model to describe these features.\cite{pw} However,
treating the Hubbard model has proven a stiff task because of the
difficulty of handling both itinerant and atomic aspects simultaneously
in a single theoretical scheme.  This awkward nature of the Hubbard model
makes ordinary analytic approach spiritless.  For this  reason, only numerical
approaches have described the doped Hubbard model in a relatively
successful way.\cite{da} These numerical works performed in two dimensions
have shown the appearance of the gap states and the shift of the chemical
potential, which agree with XAS experiments\cite{che}. There are
other theoretical works showing these two phenomena in terms of the different
models such as the Emery model and the Anderson lattice model. The former
has been studied by cluster calculation\cite{hor} and the latter by slave
boson representation\cite{mel}.  It is believed that the above phenomena
probably have nothing to do with high-$T_c$ superconductivity but with general
strong correlation effects.  Thus we expect that these strong correlation
effects could be seen in infinite dimensions, too.  This leads us to studying
the doped Hubbard model in infinite dimensions where an analytic approach is
possible. The methods developed in infinite dimensions\cite{me} have explained
the Mott-Hubbard transition\cite{ho,gk,rkz} successfully
and even quantitatively described the correlated insulator $V_2O_3$,
recently\cite{hk,ga}. Large-dimension approaches, however, have never been
applied extensively to the doped Hubbard model in the ground state showing the
gap states, the behavior of the chemical potential, and the MIR band in the
optical conductivity.

The purpose of this paper is to show that the above mentioned strong
correlation effects shown in the finite dimensions can also be seen in the
infinite dimensions. To do this, we obtain the single-particle density of
states (DOS) and the optical conductivity for the doped
Hubbard model on a Bethe lattice with infinite connectivities
through a purely analytical method. Our analytical results are
comparable with the numerical works in two dimensions\cite{da} and XAS
experiments\cite{che}.

The single-particle DOS can be written as\cite{ho}
\begin{equation}
\rho_{\sigma}(\omega)=\frac{1}{\pi N}\lim_{\epsilon\rightarrow 0^+}\sum_j
{\rm Re} \Xi_{jj}(z)|_{z=-i\omega+\epsilon},
\end{equation}
where
\begin{equation}
\Xi_{jj}(z)=\frac{1}{z-\alpha_0+\frac{\Delta_1}{z-
\alpha_1+\frac{\Delta_2}{z-\alpha_2+\ddots}}},
\end{equation}
where
$\alpha_{\nu}=(iLf_{\nu},f_{\nu})/(f_{\nu},f_{\nu})$,
$\Delta_{\nu}=(f_{\nu},f_{\nu})/(f_{\nu-1},f_{\nu-1})$. The inner product is
defined by $(A,B)=\langle\Psi_0|\{A,B^{\dagger}\}|\Psi_0\rangle$ where $A$
and $B$ are operators of the Liouville space, $B^{\dagger}$ is the adjoint of
$B$.  We choose $f_0=c_{j\sigma}$ and calculate $\alpha_{\nu}$ and
$\Delta_{\nu}$ using the recurrence relation $f_{\nu+1}=iLf_{\nu}-
\alpha_{\nu}f_{\nu}+\Delta_{\nu}f_{\nu-1}$.\cite{hong}
We apply this formalism to the Hubbard model
\begin{equation}
H=-t\sum_{<jk>\sigma}c_{j\sigma}^{\dagger}c_{k\sigma}+\frac{U}{2}\sum_
{j\sigma}n_{j\sigma}n_{j,-\sigma},
\end{equation}
where $<jk>$ means nearest  neighbor sites and $t$ and $U$ are
the hopping integral and the on-site Coulomb interaction, respectively.

Before going to the doped system, we determine the gap width of the
undoped insulator by getting the single-particle DOS. We suppose the ground
state is antiferromagnetic with quantum fluctuation $\delta$, i.e.,
$\langle n_{j\sigma}\rangle=(1-\delta),\langle n_{k\sigma}\rangle=\delta,
\langle n_{l\sigma}\rangle=(1-\delta)$, and so on.
Since a substantial amount of quantum fluctuation brings the system in
between the N\'{e}el state and the paramagnetic state, it is hard to
treat the intermediate state analytically. We know, however, that the
antiferromagnetic ordering effect mainly appears in $\alpha_{\nu}$
and the band width is mainly determined by $\Delta_{\nu}$. Thus using
the above formulae we get
\begin{equation}
\alpha_{2\nu}=-iU(1-\delta), \hspace{1cm}
\alpha_{2\nu+1}=-iU\delta
\end{equation}
for $\nu\geq 0$ and
\[ \Delta_1=U^2\delta(1-\delta)+qt^2, \hspace{0.5cm}
\Delta_2=\frac{2U^2qt^2\delta(1-\delta)+q^2t^4}{U^2\delta(1-\delta)+qt^2} \]
\begin{eqnarray}
\Delta_{2\nu+1} &= &U^2\delta(1-\delta)[1+O(\frac{t^2}{U^2\delta(1-
\delta)})]
    \equiv a, \nonumber\\
\Delta_{2\nu+2} &= &2qt^2[1+O(\frac{t^2}{U^2\delta(1-\delta)})] \equiv b,
\end{eqnarray}
for $\nu \ge 1$, where $q$ is the coordination number. To make the kinetic
energy finite as $q\rightarrow\infty$, one usually scales $t$ as
$\frac{t_*}{\sqrt{2q}}$,\cite{me} and we set $t_*=1$ in what follows. The
single-particle DOS obtained using Eqs. (4) and (5) yields the
gap of width $2\sqrt{U^2(\frac{1}{2}-\delta)^2+(\sqrt{a}-
\sqrt{b})^2}$ which corresponds to the charge transfer gap in high-$T_c$
cuprates. It is shown in Fig. 1 (a) for $U=3$ and $\delta=0.2$.

For the doped system, the antiferromagnetic order is quickly broken down upon
doping and the magnetic phase is not yet clear.  As a naive candidate we
consider the paramagnetic state which is reasonable for the
appropriately doped regime. For this state, we may write
$\langle n_{i\sigma}\rangle=\frac{(1-x)}{2}$ for all site, where $x$ is the
doping parameter.  Then we get
\begin{eqnarray}
\alpha_{4\nu} &=&-i\frac{U}{2} [1-(2\nu+1)x+O(x^3)
+O(\frac{1}{U^2})],\nonumber\\
\alpha_{4\nu+1} &=&-i\frac{U}{2} [1+(2\nu+1)x+O(x^3)
       +O(\frac{1}{U^2})],\nonumber\\
\alpha_{4\nu+2} &=&-i\frac{U}{2} [1-(2\nu+2)x+O(x^3)
       +O(\frac{1}{U^2})],\nonumber\\
\alpha_{4\nu+3} &=&-i\frac{U}{2} [1+(2\nu+2)x+O(x^3)
       +O(\frac{1}{U^2})],
\end{eqnarray}
for $\nu\geq 0$, and
\begin{eqnarray}
\Delta_1 &=&\frac{U^2}{4}(1-x^2)+\frac{1}{2}, \hspace{0.5cm} \Delta_2
=\frac{U^2(1+x^2)+1}{U^2(1-x^2)+2}, \nonumber \\
\Delta_3 &=&\frac{U^2}{4}(1-4x^2)[1+O(\frac{1}{U^2})], \hspace{0.5cm}
\Delta_4=\frac{1+3x^2}{1-3x^2}[1+O(\frac{1}{U^2})], \nonumber \\
\Delta_{4\nu+1}    &=&\frac{U^2}{4}[1-\{(2\nu+1)x\}^2+O(x^4)
         +O(\frac{1}{U^2})],\nonumber\\
\Delta_{4\nu+2}    &=&1+O(x^2)+O(\frac{1}{U^2}),\nonumber\\
\Delta_{4\nu+3}    &=&\frac{U^2}{4}[1-\{(2\nu+2)x\}^2+O(x^4)
         +O(\frac{1}{U^2})],\nonumber\\
\Delta_{4\nu+4}    &=&1+O(x^2)+O(\frac{1}{U^2}),
\end{eqnarray}
for $\nu\ge 1$.

The key approximation in getting Eqs. (4)-(7) is the Hartree-Fock
approximation which corresponds to neglecting the spatial correlations, which
is valid in the large-$q$ limit. Therefore, in calculating the ground state
correlation function, we use a decoupling scheme, e.g., $\langle
n_{j\sigma}n_{l\sigma}\rangle=\langle n_{j\sigma}\rangle \langle
n_{l\sigma}\rangle$.  The same decoupling scheme has been done on the average
of an operator containing $[H,n_{j\sigma}]$ whose average vanishes, since no
steady current exists in the ground state.

 From Eq. (7), we find a remarkable fact occurring in the doped system.
The dimension of the Liouville space of the operator $c_{j\sigma}(t)$
was infinite for the half-filled Hubbard model.\cite{ho}
The effective dimensionality for the doped Hubbard model, however,
is surprisingly reduced. One can see this in Eq. (7).
If one neglects higher order terms such as $O(x^4)$ and $O(1/U^2)$
compared with unity, which is valid for small-$x$ and large-$U$,
the Liouville space of $c_{j\sigma}(t)$ becomes $(2n-1)$-dimensional
when $x=1/n$, where $n$ is an integer except $0$. One can find vanishing
$\Delta_{\nu}$ at $x=1/n$\cite{note}, then the continued fraction (2)
is truncated and we get the single-particle DOS  as a linear combination
of the sharp Lorentzian peaks with appropriate weights. The DOS for
various hole concentrations for $U=3$ are shown in Fig. 1 (b)-(d).
It is clearly seen that the insulator gap is filled with the gap states upon
doping and the chemical potential is located just below the gap states
(low-energy peak) and near the top of the valence band of the undoped
antiferromagnetic insulator.  If we increase hole concentration, we see that
the low-energy peak increases as much as the reduced spectral weight in both
lower and upper Hubbard bands.  This low-energy peak may be caused by the
enhanced spin fluctuations at site $j$ due to increased holes around it. This
can be seen as the first peak of the two prepeaks in XAS
experiments.\cite{che} For the electron doping, there is a reflection
symmetry with hole doping because of particle-hole symmetry, so the chemical
potential lies just above the peak induced by the spin fluctuation and near
the bottom of the conduction band of the undoped insulator. We show the
relation between the position of the chemical potential $\mu$ and the
electronic density $\langle n\rangle$ for both hole doped, $\langle n \rangle
< 1$, and  electron doped, $\langle n \rangle >1$, regimes in Fig. 2. It is
clearly shown that the chemical potential eventually crosses the edges of the
insulator gap when the doping concentration exceeds a certain value. Both
experimental\cite{che} and theoretical works\cite{da,hor,mel}
including present one show that the gap states newly appear and the chemical
potential shifts across a significant portion of the insulator gap when the
doping is changed from holes to electrons.

We now show that the gap states created by doping are responsible for the
appearance of the MIR band in the optical conductivity. The optical
conductivity in infinite dimensions is obtained using the following
formula\cite{th}.
\begin{equation}
\sigma(\omega)=\sigma_0 \int d w^{\prime} \int d\epsilon A_0(\epsilon)
        A(\epsilon,\omega^{\prime}) A(\epsilon,\omega^{\prime}+\omega)
        \frac{f(\omega^{\prime})-f(\omega^{\prime}+\omega)}{\omega},
\end{equation}
where $A_0(\epsilon)=\frac{\sqrt{2-\epsilon^2}}{\pi}$
for the Bethe lattice, $f(\omega)$ is the Fermi distribution function and
$\sigma_0=\frac{\pi t^2_* e^2 a^2 N}{2 \hbar V}$ where $a, N, V$ are lattice
constant, number of lattice sites, volume, respectively, and
$A(\epsilon, \omega)=-\frac{1}{\pi} {\rm Im} G({\bf k},\omega)
=-\frac{1}{\pi}{\rm Im}[\omega+i\eta-\epsilon-\Sigma(\omega)]^{-1}$. Use of
the momentum-independence of the self-energy in infinite dimensions
has been made. We get $A(\epsilon, \omega)$ using the relations between the
on-site Green function $G(\omega)=-i\Xi_{jj}(-i\omega)$ and the self-energy
$\Sigma(\omega)$ applicable to the Bethe lattice, which are
$\Sigma(\omega)=\omega-\frac{G(\omega)}{2}-\frac{1}{G(\omega)}$ for
paramagnetic state and $\Sigma_{\sigma}(\omega)=\omega-\frac{G_
{-\sigma}(\omega)}{2}-\frac{1}{G_{\sigma}(\omega)}$ for antiferromagnetic
state.\cite{gk,hk} The optical
conductivity $\sigma(\omega)$ has been shown in units of $\sigma_0$ in
Fig. 3. The MIR band is clearly seen and quite similar to the
experiment\cite{uch} though the doping concentrations are larger than those of
experiment.  There is also a numerical work for the MIR band in two
dimensions.\cite{dago2}

It is quite interesting to compare our analytic results in
infinite dimensions with the numerical results in two
dimensions\cite{da} and XAS experiments\cite{che}. Both
theoretical results show the following agreements. The doping effect removes
some of the spectral weight from both the lower and upper Hubbard band and
creates new states inside the insulator gap. However, the chemical potential
lies near the top of the valence and the bottom of the conduction band for
hole and electron doping, respectively. It agrees with XAS experiment. In the
optical conductivity, on the other hand, the MIR band is shown in the
insulator gap upon doping.

In conclusion, all of the above features obtained analytically in infinite
dimensions have also shown by the numerical work in two dimensions, which
implies that some of the strong correlation effects may play a similar role in
the physics of both two and infinite dimensions.  Therefore, we argue that the
phenomena treated in this work are the general properties of the strongly
correlated system described by the doped Hubbard model.

 The  authors   wish  to   express  their gratitude   to  the International
Center for Theoretical Physics for  financial support and hospitality.
 One of authors (J.H.) also thanks the Government of Japan for  support.
This work  has   been supported by SNU-CTP  and   Basic  Science  Research
Institute  Program,  Ministry   of Education.

{\large Figure Captions}\\
\noindent Fig. 1: The density of states for $\frac{U}{t_*}=3$ with hole
doping, (a) $x=0$ (half-filling); (b) $x=\frac{1}{15}$; (c) $x=\frac{1}{10}$;
and (d) $x=\frac{1}{8}$.  The dotted vertical line indicates the position of
the chemical potential $\mu$. $0.1t_*$ has been used as a Lorentzian
half-width.  \\
\noindent Fig. 2: Electron density per site, $\langle n\rangle$ vs
$\mu$ for $\frac{U}{t_*}=3$.  The dashed line is to guide the eyes. \\
\noindent Fig. 3: The optical conductivities for $\frac{U}{t_*}=3$ with
different hole dopings. The solid, dash-dot, and dashed lines are
$x=\frac{1}{5}$, $\frac{1}{8}$, and $0$ with $\delta=0.2$, respectively.
The arrows indicate the MIR bands. $0.5t_*$ has been used as a Lorentzian
half-width. \\

\begin{thebibliography}{99}
\bibitem{note} The positive and negative $x$ represent hole and electron
        doping, respectively. The integer $n$ should be large to
        make $x$ small where the approximation is valid.
\end{thebibliography}
\end{document}